\documentclass[]{article}
\usepackage{parskip}
\RequirePackage{amsthm,amsmath,amsfonts,amssymb}
\RequirePackage[colorlinks,citecolor=blue,urlcolor=blue]{hyperref}
\RequirePackage{graphicx}
\RequirePackage{bm}
\usepackage[toc,page]{appendix}

\DeclareMathOperator*{\argmax}{arg\,max}

\providecommand{\keywords}[1]
{
	\small	
	\textbf{\textit{Keywords---}} #1
}

\title{Dynamic modeling of spike count data with Conway-Maxwell Poisson variability}
\author{Ganchao Wei\footnote{Corresponding author: ganchao.wei@uconn.edu}\\ \small Department of Statistics, University of Connecticut
	\and 
	Ian H. Stevenson\\ \small Department of Psychological Sciences, University of Connecticut\\ \small Department of Biomedical Engineering, University of Connecticut\\ \small Connecticut Institute for Brain and Cognitive Science, University of Connecticut }
\date{}

\begin{document}
	\maketitle
	
	\begin{abstract}
		In many areas of the brain, neural spiking activity covaries with features of the external world, such as sensory stimuli or an animal’s movement. Experimental findings suggest that the variability of neural activity changes over time and may provide information about the external world beyond the information provided by the average neural activity. To flexibly track time-varying neural response properties, here we developed a dynamic model with Conway-Maxwell Poisson (CMP) observations. The CMP distribution can flexibly describe firing patterns that are both under- and over-dispersed relative to the Poisson distribution. Here we track parameters of the CMP distribution as they vary over time. Using simulations, we show that a normal approximation can accurately track dynamics in state vectors for both the centering and shape parameters ($\lambda$ and $\nu$). We then fit our model to neural data from neurons in primary visual cortex and “place cells” in the hippocampus. We find that this method out-performs previous dynamic models based on the Poisson distribution. The dynamic CMP model provides a flexible framework for tracking time-varying non-Poisson count data and may also have applications beyond neuroscience.
	\end{abstract}

	\keywords{Neural spikes, Non-Poisson count data, Dynamic Conway-Maxwell Poisson, Laplace approximation}
	
	\section{Introduction}
	Although many models of neural activity assume that neurons respond with stable responses to external sensory stimuli or movements, there is substantial evidence that neural spiking activity changes over time due to adaptation and plasticity (\cite{Brown2001,Lesica2007}) as well as spontaneously (\cite{Rokni2007,Tomko1974}). At the same time, a neuron’s spiking responses on individual trials can be highly variable, even in the controlled settings with constant stimuli. In most previous research, trial-to-trial neural variability is assumed to be Poisson distributed. However, spike count distributions can be substantially more or less variable than Poisson (\cite{Maimon2009,Amarasingham2006,DeWeese2003,Kara2000}), and that the variability also appears to change over time, in many cases (\cite{Churchland2010,Churchland2011}). Here we introduce a dynamic model with Conway-Maxwell Poisson observations that can describe non-Poisson spike statistics and track changing response properties.
	
	Variability appears to be an increasingly important feature of neural responses and can act as a signature of decision making (\cite{Churchland2011}), movement preparation (\cite{Churchland2006}), or stimulus onset (\cite{Churchland2010}). Although systems neuroscience has a long history of studying how external variables influence mean firing rates, less is known about response variability. Neural activity changes on different timescales, and distinguishing changes in variability from changes in the mean response based on sparse, spike observations is a nontrivial statistical challenge (\cite{DeWeese1998}). Statistical tools to accurately track the sources of variability within a given experiment may be useful for understanding neural systems. There has been substantial work developing dynamic Poisson models (\cite{Brown2001,Eden2004}), as well as, other Poisson models that can account for fluctuating response properties with latent or observed variables (\cite{Czanner2008,Smith2003}). Several models of neural activity with non-Poisson observations have also been described (\cite{DeWeese2003,Gao2015,Pillow2012}), including a static model with Conway-Maxwell Poisson observations (\cite{Stevenson2016}). Each of these models, however, whether static or dynamic, assumes a fixed mean-variance relationship (i.e. fixed dispersion parameters). Here, to flexibly track how neural variability might change over time, we explicitly consider changes in both the mean and dispersion.
	
	Here we develop a dynamic GLM with Conway-Maxwell-Poisson (CMP) observations. The CMP distribution can account for both over- and under-dispersion in spike count data. To get the closed-form posterior for state vectors with CMP likelihood, we fit the model using a global Gaussian approximation (Laplace approximation). Since the state-space of the dynamic model has Markovian structure, inference is efficient with this approximation, and we estimate the process noise by maximizing the predictive likelihood. After illustrating the proposed method in simulations, we apply it to neural activity from primary visual cortex and place cells in the hippocampus. The dynamic CMP model can track changes in both the mean and variance of neural responses and outperforms previous Poisson models.
	
	\section{Methods}
	Here we consider a dynamic GLM with Conway-Maxwell Poisson (CMP) observations to describe time-varying spike counts. We first introduce the model. Although the CMP distribution allows us to flexibly model non-Poisson variability, one major challenge with using this model is that there are no closed-form posteriors for the CMP likelihood. Here, we fit the model using a global Gaussian approximation, and we discuss several additional technical challenges that arise when using the CMP distribution with a dynamic GLM. Code is available at \url{https://github.com/weigcdsb/COM_POISSON}. 
	
	\subsection{Dynamic Conway-Maxwell Poisson Model}
	A count observation $y$, such as the spike count for a neuron, is assumed to follow the CMP distribution, with parameters $\lambda$ and $\nu$. The probability mass function (pmf) of CMP distribution is:
	\begin{equation}
		P(Y=y|\lambda, \nu) = \frac{\lambda^y}{(y!)^\nu}\cdot\frac{1}{Z(\lambda, \nu)}
	\end{equation}
	, where $Z(\lambda, \nu)=\sum_{k=0}^{\infty}\frac{\lambda^k}{(k!)^\nu}$ is the normalizing constant. The shape parameter $\nu \geq 0$ controls different dispersion patterns, i.e. equi- $(\nu=1)$, over- $(0\leq\nu<1)$ or under-dispersion $(\nu > 1)$. Three common distributions occur as special cases: 1) the Poisson $(\nu=1)$, 2) the geometric $(\nu=0, \lambda<1)$, and 3) the Bernoulli $(\nu \rightarrow \infty)$.
	
	For multiple observations up to T steps, such as simultaneous spike counts from n neurons, denote the counts at time bin $t$ as $\bm{y}_t = (y_{1t},\ldots,y_{nt})'$, for $t=1,\ldots, T$. The corresponding CMP parameters at $t$ are $\bm{\lambda}_t = (\lambda_{1t},\ldots,\lambda_{nt})'$ and $\bm{\nu}_t = (\nu_{1t},\ldots,\nu_{nt})'$. Previous work has examined the CMP-GLM (\cite{Chatla2018,Sellers2010}), and here we focus on the dynamic version of this GLM. The CMP parameters at $t$ are modeled by two log-linear models, $\log\bm{\lambda}_t = \bm{X}_t\bm{\beta}_t$ and $\log\bm{\nu}_t = \bm{G}_t\bm{\gamma}_t$, with $\bm{\beta}_t \in \mathbb{R}^p$ and $\bm{\gamma}_t \in \mathbb{R}^q$, and $\bm{X}_t$ and $\bm{G}_t$ denote known predictors. Under the CMP-GLM, the parameters are static. Here, we assume that they  progress linearly with a Gaussian noise.
	
	The observations follow conditionally independent CMP distributions, given the state vector $\bm{\theta}_t = (\bm{\beta}'_t, \bm{\gamma}'_t)'$. 
	\begin{align}
		\bm{y}_t &\sim CMP(\bm{\lambda}_t, \bm{\nu}_t)\\
		\log\bm{\lambda}_t &= \bm{X}_t\bm{\beta}_t, \quad \log\bm{\nu}_t = \bm{G}_t\bm{\gamma}_t \nonumber
	\end{align}
	
	While the state vector  $\bm{\theta}_t$ evolves linearly with Gaussian noise:
	\begin{align}
		\bm{\theta}_1 &\sim N_{p+q}(\bm{\theta}_0, \bm{Q}_0)\\
		\bm{\theta}_t|\bm{\theta}_{t-1} &\sim N_{p+q}(\bm{F\theta}_{t-1}, \bm{Q}) \nonumber
	\end{align}
	Given the initial state mean $\bm{\theta}_0$, covariance $\bm{Q}_0$, linear dynamics $\bm{F}$ and process covariance $\bm{Q}$.
	
	\subsubsection{Inference by Gaussian approximation}
	To fit the model to data we need to estimate the time-varying state vector $\bm{\Theta} = (\bm{\theta}'_1,\ldots,\bm{\theta}'_T)\in\mathbb{R}^{(p+q)T}$. In this section, we first assume $\bm{F}$ and $\bm{Q}$ are known. Since the observations are CMP distributed, we cannot estimate $\bm{\Theta}$ in closed form. Instead, here we approximate it by a multivariate Gaussian distribution, $P(\bm{\Theta|Y}) \approx N_{(p+q)T}(\bm{\Theta|,\mu,\Sigma})$, with $\bm{Y} = (\bm{y}'_1,\ldots, \bm{y}'_T)'$. The parameters of this Gaussian are found by a global Laplace approximation, i.e. $\bm{\mu} = \argmax_{\bm{\Theta}}P(\bm{\Theta|Y})$ and $\bm{\Sigma} = -(\nabla\nabla_{\bm{\Theta}}\log P(\bm{\Theta|Y})|_{\bm{\Theta - \mu}})^{-1}$. The log-posterior is given by:
	\begin{align}
		\log P(\bm{\Theta|Y}) &= \sum_{t=1}^{T}l_t - \frac{1}{2}(\bm{\theta}_1 - \bm{\theta}_0)'\bm{Q}_0^{-1}(\bm{\theta}_1 - \bm{\theta}_0) - \frac{1}{2}\sum_{t=2}^{T}(\bm{\theta}_t - \bm{F\theta}_{t-1})'\bm{Q}^{-1}(\bm{\theta}_t - \bm{F\theta}_{t-1})\\
		l_t &= l(\bm{\theta}_t) = \log P(\bm{y}_t|\bm{\theta}_t) = \sum_{i=1}^{n}y_{it}\log \lambda_{it} - \nu_{it}y_{it}! - \log Z(\lambda_{it}, \nu_{it}) \nonumber
	\end{align}
	, where $l(\cdot)$is the log-likelihood. The log-posterior is concave (\cite{Gupta2014}), and the Markovian structure of the state vector dynamics makes it possible to optimize by Newton-Raphson (NR) in $\mathcal{O}(T)$ time (\cite{Paninski2010}).After the Newton update, we can further quantify the uncertainty for the CMP parameters and the underlying rates, as in Appendix \ref{appA}.
	
	There are several technical challenges involved with performing the Newton update with CMP observations. Firstly, in order to find the gradient and Hessian we need to calculate moments of $Y_{it}$ and $\log Y_{it}!$, which have no closed forms (\cite{Shmueli2005}). We can calculate these moments by truncated summation. However, when $\lambda \geq 2$ and $\nu \leq 1$,  truncated summation is computationally costly since we need many steps for accurate approximation. In this case, we approximate the moments using previous (\cite{Chatla2018,Gaunt2019}) asymptotic results as in Appendix \ref{appB}. A second challenge is that the Hessian is not robust to outliers. Outliers often result in the Hessian being close to singular or even positive-definite. See details in Appendix \ref{appC}. To ensure robustness, we use Fisher scoring where the observed information is replaced by the expected information. Finally, a third challenge is that the Newton updates take a long time to converge if the initial state estimate is far from the maximum of the posterior, especially when $T$ is large. To resolve this issue, we use a smoothing estimate with local Gaussian approximation as a “warm start”. Forward filtering for a dynamic Poisson model has been previously described in \cite{Eden2004}, and here we implement CMP filtering following the same rationale. Let $\bm{\theta}_{t|t-1} = E(\bm{\theta}_t|\bm{y}_1,\ldots,\bm{y}_{t-1})$ and $\bm{\Sigma}_{t|t-1} = Var(\bm{\theta}_t|\bm{y}_1,\ldots,\bm{y}_{t-1})$ be the mean and variance for the one-step prediction density and $\bm{\theta}_{t|t} = E(\bm{\theta}_t|\bm{y}_1,\ldots,\bm{y}_{t})$ and $\bm{\Sigma}_{t|t} = Var(\bm{\theta}_t|\bm{y}_1,\ldots,\bm{y}_t)$ be mean and variance for the posterior density, then the filtering update for step $t$ is given by
	\begin{align}
		\bm{\theta}_{t|t-1} &= \bm{F\theta}_{t-1|t-1}\\
		\bm{\Sigma}_{t|t-1} &= \bm{F\Sigma}_{t-1|t-1}\bm{F}' + \bm{Q} \nonumber\\
		\bm{\theta}_{t|t} &= \bm{\theta}_{t|t-1} + (\bm{\Sigma}_{t|t})\left[\frac{\partial l_t}{\partial \bm{\theta}_t}\right]_{\bm{\theta}_{t|t-1}} \nonumber\\
		(\bm{\Sigma}_{t|t})^{-1} &= (\bm{\Sigma}_{t|t-1})^{-1} - \left[\frac{\partial^2l_t}{\partial\bm{\theta}_t\partial\bm{\theta}'_t}\right]_{\bm{\theta}_{t|t-1}} \nonumber
	\end{align}
	Here, to again ensure robustness, we use Fisher scoring when updating the state covariance. We then find smoothed estimates using a backward pass (\cite{RAUCH1965}). Although doing smoothing is fast, the estimates can be inaccurate, especially when there are large changes in the state vector. In the forward filtering stage, the Gaussian approximation at each step $t$ is conducted locally at the recursive prior $\bm{\theta}_{t|t-1}$. This will be statistically inefficient when the recursive prior is too far away from the posterior mode, or when there is a large change in the state vector. Moreover, Fisher scoring reduces the efficiency of the smoother even further. The smoother provides reasonable initial estimates, but estimation accuracy is substantially improved by using Newton’s method to find the global Laplace approximation for the posterior.
	
	\subsubsection{Estimating process noise}
	\label{estQ}
	For the applications to neural data examined here, we assume that $\bm{F}=\bm{I}$. However, we still need to estimate the process noise $\bm{Q}$. When $n$ is small, especially when $n=1$, different $\bm{Q}$ values will have a substantial influence on estimation. One possible way to estimate. One possible way to estimate $\bm{Q}$ is to use an Expectation Maximization (EM) algorithm as in \cite{Macke2011}. However, using the Laplace approximation for $\bm{\Theta}$ during E-step breaks the usual guarantee of non-decreasing likelihoods in EM, and, hence, may lead to divergence. To avoid that, we could sample the posterior directly by MCMC. However, the lack of closed-form moments for the CMP distribution makes sampling computationally intensive. Here, to estimate $\bm{Q}$ robustly and quickly, we instead assume $\bm{Q}$ is diagonal and estimate it by maximizing the prediction likelihood in the filtering stage, as in \cite{Wei2021}. 
	
	\section{Results}
	
	\subsection{Tracking the mean and dispersion of spike counts over time}
\begin{figure}[h!]
		\centering
		\includegraphics[width=1\textwidth]{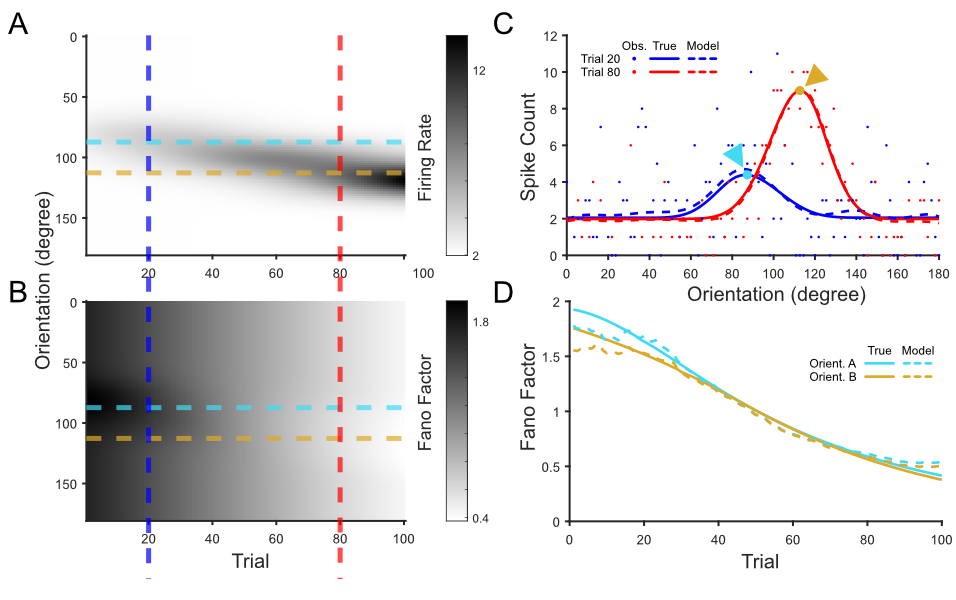}
		\caption{\textbf{A simulated neuron with a shifting firing and dispersion pattern. (A)} The tuning curve of the neuron shifts over time, with the preferred stimulus orientation changing and the response amplitude increasing. \textbf{(B)} At the same time, the variability in spiking changes from being over-dispersed relative to a Poisson distribution to under-dispersed. This leads to a decreasing Fano factor (variance-to-mean ratio) from 1.9 to 0.4 overall. \textbf{(C)} To illustrate the shifts, we show the tuning curve at two time points: Trial 20 (blue) and Trial 80 (red). Dots denote observed spike counts. The solid lines are the ground truth in mean firing rate, while the corresponding dashed lines are the fitted values. \textbf{(D)} To illustrate the shift in dispersion over time we show the true (solid) and estimated (dashed) Fano factor for two specific stimuli as a function of time. The dispersion for the early preferred orientation is shown in cyan, while the dispersion for the late preferred orientation is shown in yellow.}
		\label{fig1}
	\end{figure}

	To illustrate how the dynamic CMP model can track both time-varying mean and dispersion, we simulated a neuron with a time-varying tuning curve, where the response to 100 evenly-spaced hypothetical visual stimuli shifts over 100 trials. Here, the neuron’s tuning curve is determined by a linear combination of cubic B-spline basis functions with equally-spaced knots. The stimulus that evokes the highest average response – the “preferred orientation” – is initially ~80 deg, but shifts over the course of the experiment, and the response amplitude also increases over time (Fig. \ref{fig1}A). Meanwhile, the dispersion pattern also changes: the responses are initially over-dispersed relative to a Poisson distribution and then become under-dispersed (Fig. \ref{fig1}B). Noisy observations are sampled from the Conway-Maxwell Poisson distribution at each time (Fig. \ref{fig1}C), mimicking the types of experimental observations collected during adaptation experiments in primary visual cortex (\cite{Dragoi2000}). We then fitted the simulated spike observations using the same predictor variables as the generative model: the covariates for $\bm{\lambda_t}$ capture the tuning curve with $\bm{X}_t$ as a 10-knot cubic spline basis expansion of the orientation, and the covariate for the shape parameter $\bm{\nu_t}$ does not depend on the stimulus orientation $\bm{G}_t$. The fitted results match the ground truth well, for both the mean (Fig. \ref{fig1}C) and Fano factors (variance-to-mean ratio, Fig. \ref{fig1}D).
	
	This model-based approach provides estimates of tuning curves and dispersion at each time point. In cases where the tuning curve and variability change simultaneously, this approach can efficiently track both. By using the model with CMP observations, rather than Poisson or negative-binomial observations, the Fano factor can be both <1 (under-dispersed) and >1 (over-dispersed).
	
\begin{figure}[h!]
		\centering
		\includegraphics[width=1\textwidth]{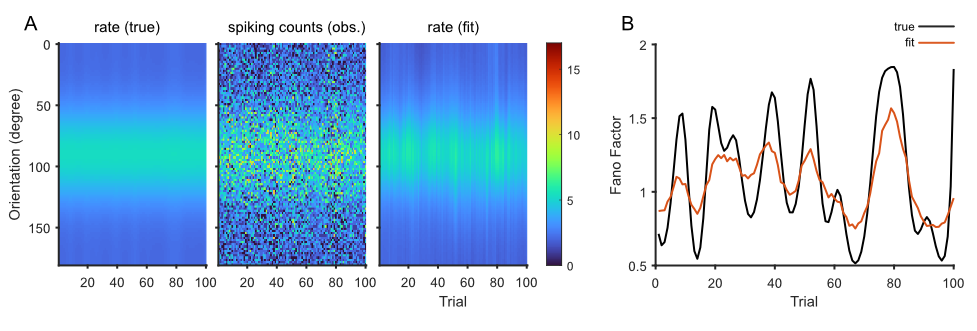}
		\caption{\textbf{Constant mean with fluctuations in dispersion. (A)} The first two panels show the true mean firing rate and the simulated observations. The last panel show the fitted mean response. \textbf{(B)} Although the mean response is constant, the Fano factor varies across the trial (black line). The colored line show the fitted result.}
		\label{fig2}
	\end{figure}

	Changes in tuning have been widely documented in systems neuroscience both due to changing environment and spontaneous nonstationarity. Changes in variability also occur, but have been less well studied. With the CMP model, the mean and dispersion are both tracked and, thus, changes in variability can occur even when the mean is stable. To illustrate this potential, we simulated a neuron whose mean firing rate is controlled to be constant, but whose Fano factor varies over time. Here $\bm{X}_t$ is a 5-knot cubic B-spline basis expansion of the orientation and $\bm{G}_t = 1$. The model recovers the true mean firing rate (Fig. \ref{fig2}A) and capture the fluctuations in variance (Fig. \ref{fig2}B) at the same time. However, the estimated Fano factor is somewhat oversmoothed when the process noise $\bm{Q}$ is optimized by maximizing the predictive likelihood (see \ref{estQ}).
	
\begin{figure}[h!]
		\centering
		\includegraphics[width=1\textwidth]{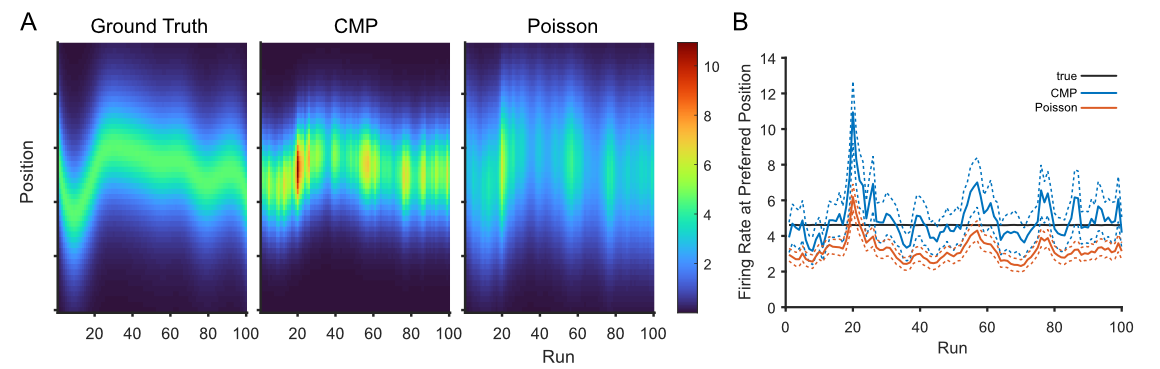}
		\caption{\textbf{Dynamic CMP and dynamic Poisson estimates differ.} Simulated over-dispersed place cell spiking is generated by the dynamic CMP model with $\nu_t = 0.1$. \textbf{(A)} We then fit dynamic CMP and dynamic Poisson models with $\bm{X}_t$, a 2-knot B-spline expansion for position ($\bm{G}_t = 1$ for the CMP). \textbf{(B)} When evaluating the response at the true preferred position for each run, the dynamic Poisson estimates are biased (under-estimated) and the uncertainty is also underestimated. The solid line gives the MAP estimates of mean firing rate, and the dashed lines show one S.D. credible intervals. The standard deviations of dynamic CMP estimates are calculated using the truncated summations (see details in APPENDIX \ref{appB}), while the standard deviations for the dynamic Poisson model are from a log-normal distribution.}
		\label{fig3}
	\end{figure}

	Although dynamic Poisson models have been applied in some neuroscientific settings, when spike counts are not Poisson distributed the model estimates can be biased. Since the dispersion influences estimates of the process noise $\bm{Q}$, estimates of the mean in the dynamic Poisson model can be effected by over- or under-dispersion. To illustrate this interaction here we simulate a place cell from the hippocampus whose “place field” drifts over time. The true mean is determined by a Gaussian function where the preferred position varies over time. The spike counts are then generated by CMP distributions, here over-dispersed with constant shape parameter $\bm{\nu}_t = 0.1$. We fit 1000 observations randomly sampled from 100 “runs” of a linear track. We find that, in this data-limited regime, the dynamic Poisson model and the dynamic CMP model give substantially different estimates of the time-varying place field (Fig. \ref{fig3}A). The dynamic Poisson model, in this case, under-estimates the firing rate at the true preferred position and under-estimates the uncertainty (Fig. \ref{fig3}B). 
	
	\subsection{Application to Experimental Data}
	We next applied our method to three publicly available datasets of extracellular spike recordings: 1) Utah array recordings of visually evoked activity from anesthetized macaque primary visual cortex (“V1 data”), 2) multi-shank silicon probe recordings from hippocampus of a rat running back-and-forth on a linear maze (“HC data”) and 3) a speed-tuned neuron recorded from the anterior pretectal nucleus (APN) of an awake mouse ("APN neuron")
	
	\subsubsection{V1 Data}
	In the V1 dataset. CRCNS pvc-11 (\cite{Kohn2016}), anesthetized macaque monkeys viewed full-field sinusoidal grating movies while neural activity was recorded by a 96-channel “Utah” array. Extracellular spiking activity was recorded on each electrode, and spike waveform segments were sorted by hand with modified competitive mixture decomposition methods (\cite{Shoham2003}). Here we use data from one animal (Monkey 1) viewing a movie of drifting sinusoidal gratings with ~100 different drift directions presented in pseudorandom order (300ms each, 30s movie in total), and the movie was repeated 120 times. Here we analyze spike counts following each stimulus presentation from the period 50-350ms after stimulus onset, considering the response delay. For further details on how the data were obtained, see \cite{Kelly2010,Smith2008}.
	
\begin{figure}[h!]
		\centering
		\includegraphics[width=1\textwidth]{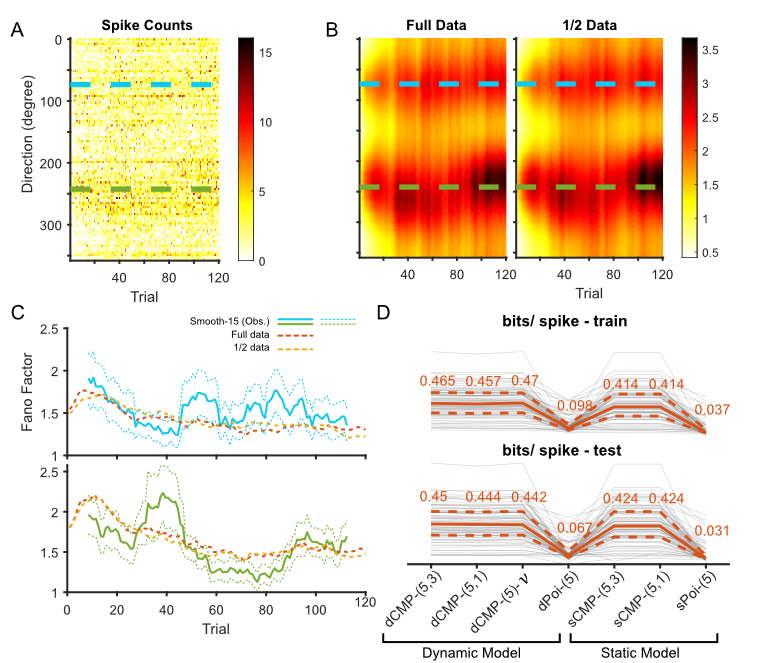}
		\caption{\textbf{Modeling nonstationary spiking from visual cortex. (A)} The spike counts of one example neuron from V1 in response to drifting grating stimuli with different drift directions presented over 120 trials. Two preferred directions (estimated by the CMP model) are marked by the dashed lines. \textbf{(B)} Estimated mean for the dynamic CMP model dCMP-(5,3) when fit to all the data and only half of the observations (held out in a speckled pattern). \textbf{(C)} Fano factor estimates for the two models, along with a direct estimate from 15-trial sliding windows, at the two preferred directions. Dashed lines denote ±1 standard deviation around the window estimates, obtained by Bayesian bootstrapping. \textbf{(D)} Model comparison for all 74 neurons in the V1 dataset. In these models, 4 are dynamic and the remaining 3 are static, with different noise distributions (Poisson vs CMP) and bases. The training and test log-likelihood ratios (bits/spike) with respect to a homogeneous static Poisson model are shown for all neurons in grey lines. The solid orange lines and numbers denote the medians, and the dashed lines show the first and third quartiles.}
		\label{fig4}
	\end{figure}

	As with many neurons in visual cortex, the responses of the neurons in this dataset are tuned to the stimulus direction. Neurons respond to some directions of stimuli more than others, but the spike counts from trial to trial are highly variable. Here, we are specifically interested in tracking changes tuning curves and changes in variability over time. Fig. \ref{fig4}A shows responses from one example neuron with a preferred direction around 240 deg. This neuron is somewhat direction insensitive, and also responds with increased spiking to stimuli moving in the opposite direction, around 70 deg. After fitting the dynamic CMP model to these data, we find that the tuning curve itself appears mostly stable, but the overall firing rate increases over the course of the recording (Fig. \ref{fig4}B). At the same time, the Fano factor decreases over the course of the recording (Fig. \ref{fig4}C). 
	
	Although the data here is structured in 120 “trials” the data are collected sequentially, and we model nonstationary at the level of individual observations. For the predictors $\bm{X}_t$ and $\bm{G}_t$ we use cubic B-spline basis functions with periodic boundary conditions over the grating directions. Results for the example neuron use 5 and 3 equally-spaced knots for $\bm{X}_t$ and $\bm{G}_t$, respectively. Fitting the model with half of the data (in a speckled hold-out pattern) gives patterns for the mean response (Fig. \ref{fig4}B) and Fano factor (Fig. \ref{fig4}C) that are similar to those using the full data. However, since the model-based approach provides a continuous estimate of the state vectors, the Fano factor estimated by the dynamic CMP model differs from a simple estimate of the Fano factor calculated using a sliding window (Fig. \ref{fig4}C).
	
	We then compare the performance of multiple models on data from all 74 neurons in this recording (Fig. \ref{fig4}D). We assess four dynamic models: (1) dynamic CMP, with 5 knots for $\bm{X}_t$ and 3 knots for $\bm{G}_t$, denoted as dCMP-(5,3); (2) dynamic CMP with $\bm{G}_t=1$, dCMP-(5,1); (3) dynamic CMP with constant $\nu_t$, dCMP-(5)-$\nu$ (fit by coordinate descent) and (4) a dynamic Poisson model, dPoi-(5). Additionally, we assess three static models: (1) static CMP, sCMP-(5,3); (2) static CMP with $\bm{G}_t=1$, sCMP-(5,1) and (3) static Poisson, sPoi-(5). The held-out log-likelihoods relative to a homogeneous static Poisson model show that the CMP-based models, both dynamic and static, outperform the Poisson-based models (Fig. \ref{fig4}D). The dynamic models perform slightly better than the corresponding static models, on average. The best performance on test data comes from modeling nonstationarity and stimulus-dependence with the full dynamic CMP model dCMP-(5,3). However, the benefit of adding nonstationary shape parameter (dCMP-(5)-$\nu$ vs. dCMP-(5,1)) and of adding stimulus-dependent shape parameter (dCMP-(5,1) vs dCMP-(5,3)) tend to be small for these data.
	
	\subsubsection{HC Data}
	In the HC dataset, CRCNS hc-3 (\cite{Mizuseki2013}), a rat was running back and forth along a 250cm linear track. Extra cellular spiking activity was recorded in dorsal hippocampus using multi-shank silicon probes. Spikes were automatically sorting using KlustaKwik followed by manual adjustment (\cite{Rossant2016}). Here we use data from one 66 min recording session (ec014-468) and analyze spike counts in 200ms bins. For further details on how the data were obtained, see \cite{Mizuseki2014}
	
\begin{figure}[h!]
		\centering
		\includegraphics[width=1\textwidth]{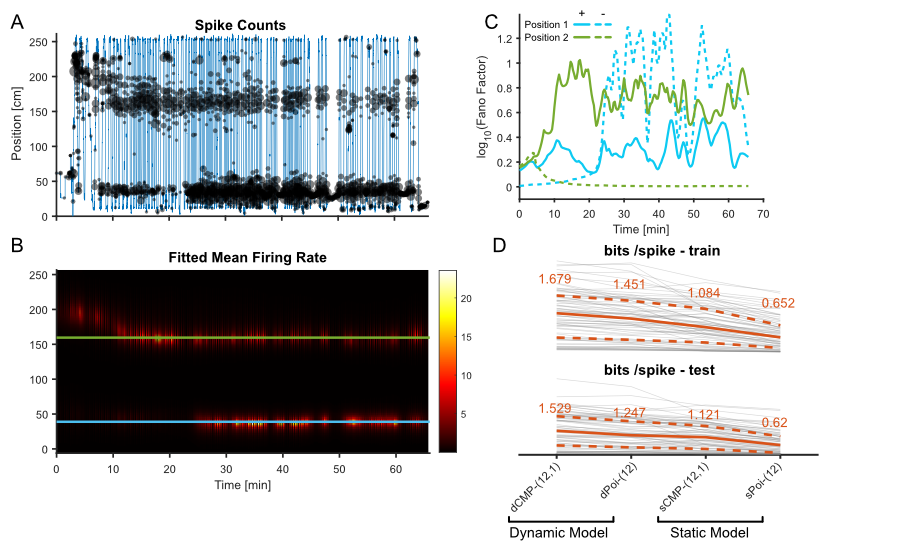}
		\caption{\textbf{Modeling nonstationarity in hippocampal spiking activity. (A)} Spike counts of an example neuron from the hippocampus recorded while a rat was running back-and-forth on a linear maze. The blue lines show the animal’s position over time. The black circles denote spike counts with the radius of each circle corresponding to number of spikes. \textbf{(B)} The heatmap shows fitted mean firing rate for the dynamic CMP model dCMP-(12,1). The colored lines show peaks for two place fields, chosen based on the model fit. \textbf{(C)} The estimated Fano factors at the two place field peaks, with each running direction (+ vs. -) shown separately. \textbf{(D)} The training and test log-likelihood ratios (bits/spike) with respect to a homogeneous static Poisson model for 78 neurons (gray lines). Here the test set log-likelihood ratios are calculated using 5\% of the data held-out in a speckled pattern. The red solid lines and numbers denote median values, while the dashed lines show the first and third quartiles.}
		\label{fig5}
	\end{figure}

	As with many neurons in hippocampus, the responses of the neurons in this dataset are tuned to the rat’s position along the track. Neurons spike at specific locations, but the place fields can also shift over time and the spike counts from run to run are highly variable. Fig. \ref{fig5}A shows an example from one neuron with two place fields where the location and firing within the place field vary over the course of the recording. Compared to the data from V1, neural responses of place cells in hippocampus tend to be sparser and more selective. Many place cells also tend to be direction tuned – spiking only when the animal is running in one direction down the track but not the other. We, thus, fit the data using a dynamic CMP model with 12 equally-spaced knots for $\bm{X}_t$ with a circular representation of position, and let $\bm{G}_t=1$.
	
	For this example neuron, the dynamic CMP model accurately tracks the time-varying place field (Fig. \ref{fig5}B). We then evaluate the fitted Fano factors at the peaks of the two place fields (Fig. \ref{fig5}C). Compared to example from the V1 data, the spiking of this example place cell is much more highly dispersed. The Fano factors vary over time and are also specific in both position and running direction. We then compare model performance on 78 neurons from this recording (19 neurons were excluded due to sparse spiking patterns). In these data, the dynamic models are generally better than the static models (Fig. \ref{fig5}D). Within groups of dynamic or static models, CMP-based models are consistently better than the Poisson-based models.
	
	\subsubsection{APN Neuron}
\begin{figure}[h!]
		\centering
		\includegraphics[width=1\textwidth]{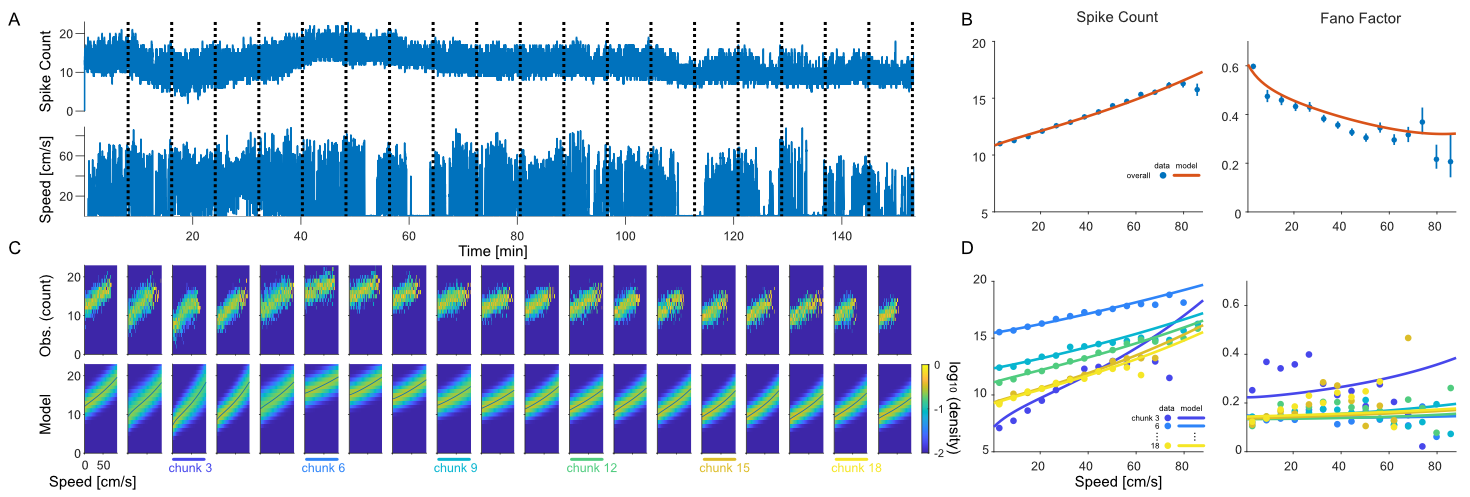}
		\caption{\textbf{Modeling a speed-tuned neuron in mouse anterior pretectal nucleus with the dynamic CMP model} Here we analyze a 160 min recording of a neuron from APN, with spike counts calculated in 200ms bins. The whole session was evenly partitioned into chunks for visualization. \textbf{(A)} Spike counts and mouse running speed for the recording session. \textbf{(B)} The spiking activity and Fano factor as a function of speed for the whole session. Dots denote results from observed spike counts. Errorbars denote 95\% confidence intervals from Bayesian bootstrapping. Red lines denote fits from the dynamic CMP model averaged over time. \textbf{(C)} The top row shows 2D histograms for spiking counts vs. running speed for each chunk. The second row shows corresponding model fitted densities (averages within each chunk). The lines show predicted average spike counts (tuning curves) for each chunk. \textbf{(D)} There is substantial variation in the tuning curves across chunks, and the Fano factor tends to be smaller on individual chunks than in the overall recording.}
		\label{fig6}
	\end{figure}

	To illustrate a case where the dynamic CMP provides a qualitatively better description of neural activity compared to previous models we show results from one neuron recorded from the Allen Institute Visual Coding Neuropixels dataset. See detailed data description in \cite{Siegle2021}. Here, when examining tuning to running speed we found a neuron in APN, whose responses were speed tuned - increasing firing with increasing running speed (Fig. \ref{fig6}A), but also highly under-dispersed relative to a Poisson distribution (Fano factor $<$1). We analyzed spike counts in 200ms bins for the 160 min recording (ecephys\_session\_id 719161530, unit\_id 950917034), and fit the spiking activity for the whole session using the dynamic CMP model. Here we use a nonlinear function of the running speed $v_t$ for the $\bm{X}_t = [1 \, f(v_t)]$ ($f(v)=v/(1+0.1v)$) and $\bm{G}_t=[1 \, v_t]$.
	
	Evaluating all of the data and averaging over time, the neuron is significantly under-dispersed (Fano factors less than 1), and both the mean response and dispersion appear tuned to running speed (Fig. \ref{fig6}B). However, we found that this neuron's speed tuning is somewhat nonstationary, with the baseline firing rate shifting over time (Fig. \ref{fig6}C). Within individual segments of the recording the Fano factor is much lower than the Fano factor evaluated for the entire recording. Moreover, the Fano factor within each chunk doesn't show a strong relationship to running speed (Fig. \ref{fig6}D). This suggests that the apparent relationship between running speed and the overall Fano factor is a byproduct of the nonstationarity. Using the static CMP model may be able to describe the underdispersion of these responses, but would miss this key feature of the data by assuming that the tuning curve is static. The dynamic Poisson model, on the other hand, is able to describe the nonstationary in the tuning curve, but  cannot describe the underdispersion of these data, since the Poisson model always assumes Fano factor = 1. 
	
	\section{Discussion}
	Here we introduced a dynamic CMP model to track changes in both the mean and dispersion of neural spikes over time. A global Laplace approximation with a smoothing-based initialization can provide accurate and computationally efficient model estimates. In both simulations and applications with experimental data we find that this model out-performs previous static and dynamic Poisson models, and may, thus, be a useful tool for understanding the role of variability in neural systems. While many studies have characterized noise (\cite{DeWeese2003,Deweese2004,Taouali2016}) or non-stationarity (\cite{Tomko1974,Wu2008}) separately, modeling changes in the mean-variance relationship directly may allow us to more closely examine the role of variability in the brain. 
	
	The extent to which the dynamic CMP model can predict neural responses more accurately than the dynamic Poisson model or static non-Poisson models depends on the neural activity itself. Here with the V1 data we found a ~6\% improvement in test log-likelihood ratios between the dynamic and static CMP, while in the HC data there is a ~35\% improvement. The extent of spontaneous changes in neural responses is somewhat unclear, some evidence suggests that neurons can be relatively stable in some circumstances (\cite{Chestek2007,Stevenson2011a,Dickey2009}), but variability appears to differ across brain areas (\cite{Mochizuki2016}). More accurate spike sorting may account for some degree of instability (\cite{Steinmetz2021}), and the degree of spontaneous changes may also depends on the brain area (\cite{Rule2019}). However, neurons do clearly change both their average responses and dispersion in many situations.
	
	Although the current model works well for fitting neural spikes, there are some potential improvements. First, the state transition matrix $\bm{F}$ is currently assumed to be known and is fixed to $\bm{I}$ for convenience in our simulations and experimental analysis. This doesn’t allow for interactions between state vectors and may limit the usage in some situations. However, when using a Laplace approximation for the state vectors, $\bm{F}$ can be estimated using the EM algorithm as in \cite{Macke2011}. Secondly, although the CMP distribution can flexibly model over- and under-dispersed data, the assumed mean-variance relationship may not be appropriate in some cases. To more flexibly model the dispersion for single observations, it may be useful to instead consider the generalized count (GC) distribution (\cite{DelCastillo2005}), which includes the CMP distribution as a special case. This model has been applied in the context of linear dynamical systems, similar to the dynamical factor analysis model,  with a fixed dispersion function (\cite{Gao2015}). To track fluctuations in dispersion more flexibly, it could be useful to allow the function to vary dynamically similar to $\nu_t$ in the dynamic CMP model here.
	
	The best modeling strategy also likely depends on what researchers want to know about the variability. Omitted variables (\cite{Stevenson2018,Goris2014}) and history effects (\cite{Uzzell2004}) can alter apparent observation noise. For example, in the hippocampus, place cell firing is highly variable on different passes through the field (\cite{Fenton1998}). This may be partially due to joint selectivity to position, speed, and head direction, as well as the influence of local field potentials. Here, rather than model these distinct covariates assuming Poisson observations, we allow the variability to be non-Poisson and introduce a dynamic GLM with CMP observations. However, doubly stochastic Poisson models (\cite{Barbieri2001}) or latent variable models with fixed mean-variance relationships (\cite{Gao2015,Pillow2012}) may also be able to account for some differences in the variance over time. Nonetheless, the dynamic CMP model may provide a useful tool for neuroscientists to study the role of variance directly. Since the static CMP model can improve decoding of external variables in some cases (\cite{Ghanbari2019}), the dynamic CMP may lead to further improvements in decoding by tracking nonstationarity in neural response properties. 
	
	\section*{Acknowledgments}
	This material is based upon work supported by the National Science Foundation under Grant No. 1931249.
	
	\bibliographystyle{unsrt}
	\bibliography{MyCollection}

\begin{thebibliography}{10}

\bibitem{Brown2001}
Emery~N. Brown, David~P. Nguyen, Loren~M. Frank, Matthew~A. Wilson, and Victor
  Solo.
\newblock {An analysis of neural receptive field plasticity by point process
  adaptive filtering}.
\newblock {\em Proceedings of the National Academy of Sciences},
  98(21):12261--12266, oct 2001.

\bibitem{Lesica2007}
Nicholas~A. Lesica, Jianzhong Jin, Chong Weng, Chun~I. Yeh, Daniel~A. Butts,
  Garrett~B. Stanley, and Jose~Manuel Alonso.
\newblock {Adaptation to Stimulus Contrast and Correlations during Natural
  Visual Stimulation}.
\newblock {\em Neuron}, 55(3):479--491, 2007.

\bibitem{Rokni2007}
U~Rokni, A~G Richardson, E~Bizzi, and H~S Seung.
\newblock {Motor learning with unstable neural representations}.
\newblock {\em Neuron}, 54(4):653--666, 2007.

\bibitem{Tomko1974}
George~J Tomko and Donald~R Crapper.
\newblock {Neuronal variability: non-stationary responses to identical visual
  stimuli}.
\newblock {\em Brain Research}, 79(3):405--418, 1974.

\bibitem{Maimon2009}
Gaby Maimon and John~a. Assad.
\newblock {Beyond Poisson: Increased Spike-Time Regularity across Primate
  Parietal Cortex}.
\newblock {\em Neuron}, 62(3):426--440, 2009.

\bibitem{Amarasingham2006}
Asohan Amarasingham, Ting~Li Chen, Stuart Geman, Matthew~T. Harrison, and
  David~L. Sheinberg.
\newblock {Spike Count Reliability and the Poisson Hypothesis}.
\newblock {\em Journal of Neuroscience}, 26(3):801--809, jan 2006.

\bibitem{DeWeese2003}
Michael~R DeWeese, Michael Wehr, and Anthony~M Zador.
\newblock {Binary spiking in auditory cortex.}
\newblock {\em The Journal of neuroscience : the official journal of the
  Society for Neuroscience}, 23(21):7940--7949, 2003.

\bibitem{Kara2000}
P~Kara, P~Reinagel, and R~C Reid.
\newblock {Low response variability in simultaneously recorded retinal,
  thalamic, and cortical neurons.}
\newblock {\em Neuron}, 27(3):635--646, 2000.

\bibitem{Churchland2010}
Mark~M Churchland, Byron~M Yu, John~P Cunningham, Leo~P Sugrue, Marlene~R
  Cohen, Greg~S Corrado, William~T Newsome, Andrew~M Clark, Paymon Hosseini,
  Benjamin~B Scott, David~C Bradley, Matthew~a Smith, Adam Kohn, J~Anthony
  Movshon, Katherine~M Armstrong, Tirin Moore, Steve~W Chang, Lawrence~H
  Snyder, Stephen~G Lisberger, Nicholas~J Priebe, Ian~M Finn, David Ferster,
  Stephen~I Ryu, Gopal Santhanam, Maneesh Sahani, and Krishna~V Shenoy.
\newblock {Stimulus onset quenches neural variability: A widespread cortical
  phenomenon}.
\newblock {\em Nature Neuroscience}, 13(3):369--378, 2010.

\bibitem{Churchland2011}
Anne~K Churchland, R.~Kiani, R.~Chaudhuri, Xiao~Jing Wang, Alexandre Pouget,
  and M.~N. Shadlen.
\newblock {Variance as a Signature of Neural Computations during Decision
  Making}.
\newblock {\em Neuron}, 69(4):818--831, 2011.

\bibitem{Churchland2006}
M~M Churchland, B~M Yu, S~I Ryu, G~Santhanam, and K~V Shenoy.
\newblock {Neural Variability in Premotor Cortex Provides a Signature of Motor
  Preparation}.
\newblock {\em Journal of Neuroscience}, 26(14):3697, 2006.

\bibitem{DeWeese1998}
Michael DeWeese and Anthony Zador.
\newblock {Asymmetric Dynamics in Optimal Variance Adaptation}.
\newblock {\em Neural Computation}, 10(5):1179--1202, 1998.

\bibitem{Eden2004}
Uri~T Eden, Loren~M. Frank, Riccardo Barbieri, Victor Solo, and Emery~N. Brown.
\newblock {Dynamic Analysis of Neural Encoding by Point Process Adaptive
  Filtering}.
\newblock {\em Neural Computation}, 16(5):971--998, may 2004.

\bibitem{Czanner2008}
Gabriela Czanner, Uri~T Eden, Sylvia Wirth, Marianna Yanike, Wendy~A Suzuki,
  and Emery~N Brown.
\newblock {Analysis of between-trial and within-trial neural spiking dynamics}.
\newblock {\em Journal of Neurophysiology}, 99(5):2672--2693, 2008.

\bibitem{Smith2003}
Anne~C. Smith and Emery~N. Brown.
\newblock {Estimating a state-space model from point process observations}.
\newblock {\em Neural Computation}, 15(5):965--991, may 2003.

\bibitem{Gao2015}
Yuanjun Gao, Lars Buesing, Krishna~V Shenoy, and John~P Cunningham.
\newblock {High-dimensional neural spike train analysis with generalized count
  linear dynamical systems}.
\newblock Technical report, 2015.

\bibitem{Pillow2012}
Jonathan~W. Pillow and James~G Scott.
\newblock {Fully Bayesian inference for neural models with negative-binomial
  spiking}.
\newblock Technical report, 2012.

\bibitem{Stevenson2016}
Ian~H. Stevenson.
\newblock {Flexible models for spike count data with both over- and under-
  dispersion}.
\newblock {\em Journal of Computational Neuroscience}, 41(1):29--43, aug 2016.

\bibitem{Chatla2018}
Suneel~Babu Chatla and Galit Shmueli.
\newblock {Efficient estimation of COM–Poisson regression and a generalized
  additive model}.
\newblock {\em Computational Statistics and Data Analysis}, 121:71--88, may
  2018.

\bibitem{Sellers2010}
Kimberly~F. Sellers and Galit Shmueli.
\newblock {A flexible regression model for count data}.
\newblock {\em The Annals of Applied Statistics}, 4(2):943 -- 961, 2010.

\bibitem{Gupta2014}
Ramesh~C. Gupta, S.~Z. Sim, and S.~H. Ong.
\newblock {Analysis of discrete data by Conway–Maxwell Poisson distribution}.
\newblock {\em AStA Advances in Statistical Analysis}, 98(4):327--343, oct
  2014.

\bibitem{Paninski2010}
Liam Paninski, Yashar Ahmadian, Daniel~Gil Ferreira, Shinsuke Koyama, Kamiar
  Rahnama~Rad, Michael Vidne, Joshua Vogelstein, and Wei Wu.
\newblock A new look at state-space models for neural data.
\newblock {\em J. Comput. Neurosci.}, 29(1–2):107–126, aug 2010.

\bibitem{Shmueli2005}
Galit Shmueli, Thomas~P. Minka, Joseph~B. Kadane, Sharad Borle, and Peter
  Boatwright.
\newblock {A useful distribution for fitting discrete data: revival of the
  Conway–Maxwell–Poisson distribution}.
\newblock {\em Journal of the Royal Statistical Society: Series C (Applied
  Statistics)}, 54(1):127--142, jan 2005.

\bibitem{Gaunt2019}
Robert~E Gaunt, Satish Iyengar, Adri~B {Olde Daalhuis}, Burcin Simsek, and B~E
  {Robert Gaunt}.
\newblock {An asymptotic expansion for the normalizing constant of the
  Conway-Maxwell-Poisson distribution}.
\newblock {\em Ann Inst Stat Math}, 71:163--180, 2019.

\bibitem{RAUCH1965}
H~E Rauch, F~Tung, and C~T Striebel.
\newblock {Maximum likelihood estimates of linear dynamic systems}.
\newblock {\em AIAA Journal}, 3(8):1445--1450, aug 1965.

\bibitem{Macke2011}
Jakob~H. Macke, Lars Buesing, John~P. Cunningham, Byron~M. Yu, Krishna~V.
  Shenoy, and Maneesh Sahani.
\newblock {Empirical models of spiking in neural populations}.
\newblock {\em Advances in Neural Information Processing Systems}, 24, 2011.

\bibitem{Wei2021}
Ganchao Wei and Ian~H. Stevenson.
\newblock {Tracking Fast and Slow Changes in Synaptic Weights From
  Simultaneously Observed Pre- and Postsynaptic Spiking}.
\newblock {\em Neural computation}, 33(10):2682--2709, sep 2021.

\bibitem{Dragoi2000}
Valentin Dragoi, Jitendra Sharma, and Mriganka Sur.
\newblock {Adaptation-Induced Plasticity of Orientation Tuning in Adult Visual
  Cortex}.
\newblock {\em Neuron}, 28(1):287--298, oct 2000.

\bibitem{Kohn2016}
A.~Kohn and M.A. Smith.
\newblock {Utah array extracellular recordings of spontaneous and visually
  evoked activity from anesthetized macaque primary visual cortex (V1)}, 2016.

\bibitem{Shoham2003}
Shy Shoham, Matthew~R. Fellows, and Richard~A. Normann.
\newblock {Robust, automatic spike sorting using mixtures of multivariate
  t-distributions}.
\newblock {\em Journal of neuroscience methods}, 127(2):111--122, aug 2003.

\bibitem{Kelly2010}
Ryan~C. Kelly, Matthew~A. Smith, Robert~E. Kass, and Tai~Sing Lee.
\newblock {Local field potentials indicate network state and account for
  neuronal response variability}.
\newblock {\em Journal of Computational Neuroscience}, 29(3):567--579, dec
  2010.

\bibitem{Smith2008}
Matthew~A. Smith and Adam Kohn.
\newblock {Spatial and temporal scales of neuronal correlation in primary
  visual cortex}.
\newblock {\em Journal of Neuroscience}, 28(48):12591--12603, nov 2008.

\bibitem{Mizuseki2013}
K.~Mizuseki, A.~Sirota, E.~Pastalkova, K.~Diba, and G.~Buzs{\'{a}}ki.
\newblock {Multiple single unit recordings from different rat hippocampal and
  entorhinal regions while the animals were performing multiple behavioral
  tasks}.
\newblock {\em CRCNS.org}, 2013.

\bibitem{Rossant2016}
Cyrille Rossant, Shabnam~N. Kadir, Dan~F.M. Goodman, John Schulman,
  Maximilian~L.D. Hunter, Aman~B. Saleem, Andres Grosmark, Mariano Belluscio,
  George~H. Denfield, Alexander~S. Ecker, Andreas~S. Tolias, Samuel Solomon,
  Gy{\"{o}}rgy Buzski, Matteo Carandini, and Kenneth~D. Harris.
\newblock {Spike sorting for large, dense electrode arrays}.
\newblock {\em Nature Neuroscience 2016 19:4}, 19(4):634--641, mar 2016.

\bibitem{Mizuseki2014}
Kenji Mizuseki, Kamran Diba, Eva Pastalkova, Jeff Teeters, Anton Sirota, and
  Gy{\"{o}}rgy Buzs{\'{a}}ki.
\newblock {Neurosharing: large-scale data sets (spike, LFP) recorded from the
  hippocampal-entorhinal system in behaving rats}.
\newblock {\em F1000Research 2014 3:98}, 3:98, jul 2014.

\bibitem{Siegle2021}
Joshua~H. Siegle, Xiaoxuan Jia, S{\'{e}}verine Durand, et~al.
\newblock {Survey of spiking in the mouse visual system reveals functional
  hierarchy}.
\newblock {\em Nature 2021 592:7852}, 592(7852):86--92, jan 2021.

\bibitem{Deweese2004}
Michael~R Deweese and Anthony~M Zador.
\newblock {Shared and private variability in the auditory cortex.}
\newblock {\em Journal of neurophysiology}, 92(3):1840--1855, 2004.

\bibitem{Taouali2016}
Wahiba Taouali, Giacomo Benvenuti, Pascal Wallisch, Fr{\'{e}}d{\'{e}}ric
  Chavane, and Laurent~U Perrinet.
\newblock {Testing the odds of inherent vs. observed overdispersion in neural
  spike counts.}
\newblock {\em Journal of neurophysiology}, 115(1):434--44, jan 2016.

\bibitem{Wu2008}
W~Wu and N~G Hatsopoulos.
\newblock {Real-time decoding of nonstationary neural activity in motor
  cortex}.
\newblock {\em Neural Systems and Rehabilitation Engineering, IEEE Transactions
  on}, 16(3):213--222, 2008.

\bibitem{Chestek2007}
Cynthia~A Chestek, Aaron~P Batista, Gopal Santhanam, Byron~M Yu, Afsheen
  Afshar, John~P Cunningham, Vikash Gilja, Stephen~I Ryu, Mark~M Churchland,
  and Krishna~V Shenoy.
\newblock {Single-neuron stability during repeated reaching in macaque premotor
  cortex}.
\newblock {\em Journal of Neuroscience}, 27(40):10742--10750, 2007.

\bibitem{Stevenson2011a}
Ian~H. Stevenson, Anil Cherian, Brian~M. London, Nicholas~A. Sachs, Eric
  Lindberg, Jacob Reimer, Marc~W. Slutzky, Nicholas~G. Hatsopoulos, Lee~E.
  Miller, and Konrad~P. Kording.
\newblock {Statistical assessment of the stability of neural movement
  representations}.
\newblock {\em Journal of Neurophysiology}, 106(2):764--774, aug 2011.

\bibitem{Dickey2009}
Adam~S Dickey, Aaron Suminski, Yali Amit, and Nicholas~G Hatsopoulos.
\newblock {Single-unit stability using chronically implanted multielectrode
  arrays}.
\newblock {\em Journal of Neurophysiology}, 102(2):1331--1339, 2009.

\bibitem{Mochizuki2016}
Yasuhiro Mochizuki, Tomokatsu Onaga, Hideaki Shimazaki, Takeaki Shimokawa,
  Yasuhiro Tsubo, Rie Kimura, Akiko Saiki, Yutaka Sakai, Yoshikazu Isomura,
  Shigeyoshi Fujisawa, Ken~Ichi Shibata, Daichi Hirai, Takahiro Furuta, Takeshi
  Kaneko, Susumu Takahashi, Tomoaki Nakazono, Seiya Ishino, Yoshio Sakurai,
  Takashi Kitsukawa, Jong~Won Lee, Hyunjung Lee, Min~Whan Jung, Cecilia Babul,
  Pedro~E. Maldonado, Kazutaka Takahashi, Fritzie~I. Arce-McShane, Callum~F.
  Ross, Barry~J. Sessle, Nicholas~G. Hatsopoulos, Thomas Brochier, Alexa
  Riehle, Paul Chorley, Sonja Gr{\"{u}}n, Hisao Nishijo, Satoe Ichihara-Takeda,
  Shintaro Funahashi, Keisetsu Shima, Hajime Mushiake, Yukako Yamane, Hiroshi
  Tamura, Ichiro Fujita, Naoko Inaba, Kenji Kawano, Sergei Kurkin, Kikuro
  Fukushima, Kiyoshi Kurata, Masato Taira, Ken~Ichiro Tsutsui, Tadashi Ogawa,
  Hidehiko Komatsu, Kowa Koida, Keisuke Toyama, Barry~J. Richmond, and Shigeru
  Shinomoto.
\newblock {Similarity in Neuronal Firing Regimes across Mammalian Species}.
\newblock {\em Journal of Neuroscience}, 36(21):5736--5747, may 2016.

\bibitem{Steinmetz2021}
Nicholas~A. Steinmetz, Cagatay Aydin, Anna Lebedeva, Michael Okun, Marius
  Pachitariu, Marius Bauza, Maxime Beau, Jai Bhagat, Claudia B{\"{o}}hm,
  Martijn Broux, Susu Chen, Jennifer Colonell, Richard~J. Gardner, Bill Karsh,
  Fabian Kloosterman, Dimitar Kostadinov, Carolina Mora-Lopez, John
  O'Callaghan, Junchol Park, Jan Putzeys, Britton Sauerbrei, Rik~J.J. van Daal,
  Abraham~Z. Vollan, Shiwei Wang, Marleen Welkenhuysen, Zhiwen Ye, Joshua~T.
  Dudman, Barundeb Dutta, Adam~W. Hantman, Kenneth~D. Harris, Albert~K. Lee,
  Edvard~I. Moser, John O'Keefe, Alfonso Renart, Karel Svoboda, Michael
  H{\"{a}}usser, Sebastian Haesler, Matteo Carandini, and Timothy~D. Harris.
\newblock {Neuropixels 2.0: A miniaturized high-density probe for stable,
  long-term brain recordings}.
\newblock {\em Science}, 372(6539), apr 2021.

\bibitem{Rule2019}
Michael~E. Rule, Timothy O'Leary, and Christopher~D. Harvey.
\newblock {Causes and consequences of representational drift}.
\newblock {\em Current Opinion in Neurobiology}, 58:141--147, oct 2019.

\bibitem{DelCastillo2005}
Joan del Castillo and Marta P{\'{e}}rez-Casany.
\newblock {Overdispersed and underdispersed Poisson generalizations}.
\newblock {\em Journal of Statistical Planning and Inference}, 134(2):486--500,
  oct 2005.

\bibitem{Stevenson2018}
Ian~H. Stevenson.
\newblock {Omitted Variable Bias in GLMs of Neural Spiking Activity}.
\newblock {\em Neural Computation}, 30(12):3227--3258, oct 2018.

\bibitem{Goris2014}
Robbe~L.T. Goris, J.~Anthony Movshon, and Eero~P. Simoncelli.
\newblock {Partitioning neuronal variability}.
\newblock {\em Nature Neuroscience 2014 17:6}, 17(6):858--865, apr 2014.

\bibitem{Uzzell2004}
V.~J. Uzzell and E.~J. Chichilnisky.
\newblock {Precision of spike trains in primate retinal ganglion cells}.
\newblock {\em Journal of Neurophysiology}, 92(2):780--789, aug 2004.

\bibitem{Fenton1998}
Andr{\'{e}}~A. Fenton and Robert~U. Muller.
\newblock {Place cell discharge is extremely variable during individual passes
  of the rat through the firing field}.
\newblock {\em Proceedings of the National Academy of Sciences of the United
  States of America}, 95(6):3182--3187, 1998.

\bibitem{Barbieri2001}
Riccardo Barbieri, Michael~C Quirk, Loren~M Frank, Matthew~A Wilson, and
  Emery~N Brown.
\newblock {Construction and analysis of non-Poisson stimulus-response models of
  neural spiking activity}.
\newblock {\em Journal of Neuroscience Methods}, 105(1):25--37, jan 2001.

\bibitem{Ghanbari2019}
Abed Ghanbari, Christopher~M. Lee, Heather~L. Read, and Ian~H. Stevenson.
\newblock {Modeling stimulus-dependent variability improves decoding of
  population neural responses}.
\newblock {\em Journal of Neural Engineering}, 16(6), 2019.

\end{thebibliography}
	
	\begin{appendices}
		\section{Quantifying Uncertainties}\label{appA}
		
		After convergence, we have an approximation of the log-posterior $P(\bm{\theta}_t|\bm{Y}) \approx N(\bm{\theta}_t|\bm{\mu}_t, \bm{\Sigma}_t)$, and we can use this approximation to quantify the uncertainty about the CMP parameters, as well as about the mean rate at each time.
		
		The CMP parameters are log-normal distributed. Let $\bm{Z}_{it} = \begin{pmatrix}
			\bm{x}'_{it} & \bm{0} \\
			\bm{0} & \bm{g}'_{it} 
		\end{pmatrix}$, then $(\lambda_{it},\nu_{it})'=\exp(\bm{Z}_{it}\bm{\theta}_t) \sim \text{Lognormal}_2(\bm{Z}_{it}\bm{\mu}_t, \bm{Z}_{it}\bm{\Sigma}_t\bm{Z}'_{it})$. Denote the variance of CMP parameters as $\bm{V}_{it}$., where for $\bm{a} = \bm{Z\mu}$ and $\bm{S} = \bm{Z\Sigma Z'}$, $[V]_{mn} = e^{a_m + a_n + \frac{1}{2}(S_{mm} + S_{nn})}(e^{S_{mn}}-1)$
		
		The conditional mean firing rate is $\delta_{it} = E(Y_{it})$, whose variance can be calculated by the Delta method:
		\begin{align}
			\widehat{Var}(\delta_{it}) &=  \begin{pmatrix}
				\frac{\partial\delta_{it}}{\partial\lambda_{it}} & \frac{\partial\delta_{it}}{\partial\nu_{it}} 
			\end{pmatrix}\bm{V}_{it}\begin{pmatrix}
				\frac{\partial\delta_{it}}{\partial\lambda_{it}} \\ \frac{\partial\delta_{it}}{\partial\nu_{it}}
			\end{pmatrix}\\
			\frac{\partial\delta_{it}}{\partial\lambda_{it}} &= \frac{\partial^2\log Z_{it}}{\partial\log\lambda_{it}\partial\lambda_{it}} = \frac{Var(Y_{it})}{\lambda_{it}}\\
			\frac{\partial\delta_{it}}{\partial\nu_{it}} &= \frac{\partial^2\log Z_{it}}{\partial\log\lambda_{it}\partial\nu_{it}} = -Cov(Y_{it}, \log Y_{it}!)
		\end{align}
		We can calculate the moments as in Appendix \ref{appB}, or we can use simpler approximations $E(Y) = \lambda^{1/\nu} - \frac{\nu-1}{2\nu}$ when $\nu\leq 1$ or $\lambda > 10^\nu$. Then $\frac{\partial\delta_{it}}{\partial\lambda_{it}}\approx\frac{1}{\nu_{it}}\lambda_{it}^{1/\nu_{it} - 1}$ and $\frac{\delta_{it}}{\nu_{it}} \approx -\frac{\lambda_{it}^{1/\nu_{it}}\log \lambda_{it}}{\nu_{it}^2} - \frac{1}{2\nu_{it}^2}$.
		
		\section{Moments approximation for Conway-Maxwell Poisson distribution}\label{appB}
		To estimate the state-vector for the dynamic CMP model, we need to find first and second moments for $Y$ and $\log Y!$. For $Y\sim CMP(\lambda, \nu)$,
		\begin{align}
			Z(\lambda, \nu) &= \sum_{k=0}^{\infty}\frac{\lambda^k}{(k!)^\nu}\\
			E(Y) &= \frac{\partial\log Z}{\partial \log\lambda} = \frac{1}{Z}\sum_{k=0}^{\infty}\frac{k\lambda^k}{(k!)^\nu} \nonumber\\
			Var(Y) &= \frac{\partial^2\log Z}{\partial(\log \lambda)^2} = \frac{1}{Z}\sum_{k=0}^{\infty}\frac{k^2\lambda^k}{(k!)^\nu} - E^2(Y) \nonumber\\
			E(\log Y!) &= -\frac{\partial\log Z}{\partial\nu} = \frac{1}{Z}\sum_{k=0}^{\infty}\frac{(\log k!)\lambda^k}{(k!)^\nu} \nonumber\\
			Var(\log Y!) &= \frac{\partial^2\log Z}{\partial \nu^2} = \frac{1}{Z}\sum_{k=0}^{\infty}\frac{(\log k!)^2\lambda^k}{(k!)^\nu} - E^2(\log Y!) \nonumber\\
			Cov(Y, \log Y!) &= -\frac{\partial^2\log Z}{\partial \log\lambda\partial\nu} = \frac{1}{Z}\sum_{k=0}^{\infty}\frac{(\log k!) k\lambda^k}{(k!)^\nu} - E(\log Y!)E(Y) \nonumber
		\end{align}
		Generally, these moments can be calculated by truncated summation.
		
		However, when $\lambda \geq 2$ and $\nu \leq 1$, we need many steps for accurate approximation. In this case, we make use of a previous asymptotic results (\cite{Chatla2018,Gaunt2019}) for efficient calculation. Let $\alpha = \lambda^{1/\nu}, c_1 = \frac{\nu^2-1}{24}$ and $c_2 = \frac{\nu^2 - 1}{48} + \frac{c_1^2}{2}$,
		
		\begin{equation}
			Z(\lambda, \nu)=\frac{e^{\nu\alpha}}{\lambda^{\frac{\nu-1}{2\nu}}(2\pi)^\frac{\nu-1}{2}\sqrt{\nu}}(1+c_1(\nu\alpha)^{-1} + c_2(\nu\alpha)^{-2} + \mathcal{O}(\lambda^{-3/\nu}))
		\end{equation}
		Then the moments can be calculated as follows:
		\begin{align}
			E(Y) &= \alpha - \frac{\nu-1}{2\nu} - \frac{\nu^2-1}{24\nu^2}\alpha^{-1}-\frac{\nu^2-1}{24\nu^3}\alpha^{-2} + \mathcal{O}(\alpha^{-3})\\
			Var(Y) &= \frac{\alpha}{\nu} + \frac{\nu^2-1}{24\nu^3}\alpha^{-1} + \frac{\nu^2-1}{12\nu^4}\aleph^{-2} + \mathcal{O}(\alpha^{-3}) \nonumber\\
			E(\log Y!) &= \alpha\left(\frac{\log\lambda}{\nu} - 1\right) + \frac{\log\lambda}{2\nu^2} + \frac{1}{2\nu} + \frac{\log 2\pi}{2} \nonumber \\
			&- \frac{\alpha^{-1}}{24}\left(1 + \frac{1}{\nu^2} + \frac{\log\lambda}{\nu} - \frac{\log\lambda}{\nu^3}\right) \nonumber\\
			&- \frac{\alpha^{-2}}{24}\left(\frac{1}{\nu^3} + \frac{\log\lambda}{\nu^2} - \frac{\log \lambda}{\nu^4}\right) + \mathcal{O}(\alpha^{-3}) \nonumber \\
			Var(\log Y!) &= \frac{\alpha(\log\lambda)^2}{\nu^3} + \frac{\log\lambda}{\nu^3} + \frac{1}{2\nu^2} \nonumber\\
			&+ \frac{\alpha^{-1}}{24\nu^5}[-2\nu^2 + 4\nu\log\lambda + (-1 + \nu^2)(\log\lambda)^2] \nonumber\\
			&+ \frac{\alpha^{-2}}{24\nu^6}[-3\nu^2 - 2\nu(-3 + \nu^2)\log\lambda + 2(-1 + \nu^2)(\log\lambda)^2] + \mathcal{O}(\alpha^{-3}) \nonumber \\
			Cov(Y, \log Y!) &= \frac{\alpha\log\lambda}{\nu^2} + \frac{1}{2\nu^2} + \frac{\alpha^{-1}}{24}\left(\frac{2}{\nu^3} + \frac{\log\lambda}{\nu^2} - \frac{\log\lambda}{\nu^4}\right) \nonumber\\
			&-\frac{1}{24\alpha^2}\left(\frac{1}{\nu^2} - \frac{3}{\nu^4} - \frac{2\log\lambda}{\nu^3} + \frac{2\log\lambda}{\nu^5}\right) + \mathcal{O}(\alpha^{-3}) \nonumber
		\end{align}

		\section{Gradient and Hessian of the log-posterior}\label{appC}
		We estimate the state vector by maximizing the log-posterior with Newton-Raphson updates. Denote $f = P(\bm{\Theta|Y})$, the $(k+1)$-th update of NR algorithm is $\bm{\Theta}^{(k+1)} = \bm{\Theta}^{(k)} + [\nabla\nabla_{\bm{\Theta}^{(k)}}f]^{-1}\nabla_{\bm{\Theta}^{(k)}}f$. 
		The gradient is:
		\begin{align}
			\nabla_{\bm{\Theta}}f &= \left[\left(\frac{\partial f}{\partial \bm{\theta}_1}\right)',\ldots,\left(\frac{\partial f}{\partial \bm{\theta}_T}\right)'\right]'\\
			\frac{\partial f}{\partial\bm{\theta}_1} &= \frac{\partial l_1}{\partial\bm{\theta}_1} -  \bm{Q}_0^{-1}(\bm{\theta}_1 - \bm{\theta}_0)   +  \bm{F'Q}^{-1}(\bm{\theta}_2 - \bm{F\theta}_{1}) \nonumber\\
			\frac{\partial f}{\partial\bm{\theta}_t} &= \frac{\partial l_t}{\partial\bm{\theta}_t} -  \bm{Q}^{-1}(\bm{\theta}_t - \bm{F\theta}_{t-1})   +  \bm{F'Q}^{-1}(\bm{\theta}_{t+1} - \bm{F\theta}_{t}) \nonumber\\
			\frac{\partial f}{\partial\bm{\theta}_T} &= \frac{\partial l_T}{\partial\bm{\theta}_T} -  \bm{Q}^{-1}(\bm{\theta}_T - \bm{F\theta}_{T-1})    \nonumber\\
			\frac{\partial l_t}{\partial\bm{\theta}_t} &= \sum_{i = 1}^{n}\begin{pmatrix}
				{\left( y_{it} - E\left( Y_{it} \right) \right)\bm{x}}_{it} \\
				{\nu_{it}\left( E\left( \log{Y_{it}!} \right) - \log{y_{it}!} \right)\bm{g}}_{it} \\
			\end{pmatrix} \nonumber
		\end{align}
		
		The Hessian is:
		\begin{align}
			\nabla\nabla_{\bm{\Theta}}f &=
			\begin{pmatrix}
				\frac{\partial^{2}f}{\partial\bm{\theta}_{1}\partial\bm{\theta}_{1}^{'}} & \bm{F}'\bm{Q}^{-1} &\bm{0} & \cdots & \bm{0} \\
				\bm{Q}^{- 1}\bm{F} & \frac{\partial^{2}f}{\partial\bm{\theta}_{2}\partial\bm{\theta}_{2}^{'}}&\bm{F}'\bm{Q}^{-1} & \cdots & \vdots \\
				\bm{0} & \bm{Q}^{-1}\bm{F}  & \frac{\partial^{2}f}{\partial\bm{\theta}_{3}\partial\bm{\theta}_{3}^{'}} & \cdots & \vdots\\
				\vdots & \vdots & \vdots & \ddots & \vdots \\
				\bm{0} & \cdots & \cdots & \cdots & \frac{\partial^{2}f}{\partial\bm{\theta}_{T}\partial\bm{\theta}_{T}^{'}}
			\end{pmatrix}\\
			\frac{\partial^{2}f}{\partial\bm{\theta}_{1}\partial\bm{\theta}_{1}^{'}}  &= \frac{\partial^{2}l_1}{\partial\bm{\theta}_{1}\partial\bm{\theta}_{1}^{'}} - \bm{Q}_0^{-1} - \bm{F}'\bm{Q}^{-1}\bm{F} \nonumber\\
			\frac{\partial^{2}f}{\partial\bm{\theta}_{t}\partial\bm{\theta}_{t}^{'}}  &= \frac{\partial^{2}l_t}{\partial\bm{\theta}_{t}\partial\bm{\theta}_{t}^{'}} - \bm{Q}^{-1} - \bm{F}'\bm{Q}^{-1}\bm{F} \nonumber\\
			\frac{\partial^{2}f}{\partial\bm{\theta}_{T}\partial\bm{\theta}_{T}^{'}}  &= \frac{\partial^{2}l_T}{\partial\bm{\theta}_{T}\partial\bm{\theta}_{T}^{'}} - \bm{Q}^{-1} \nonumber
		\end{align}
		, where
		\begin{align}
			\frac{\partial^{2}l_t}{\partial\bm{\theta}_{t}\partial\bm{\theta}_{t}^{'}} &= \sum_{i = 1}^{n}\begin{pmatrix}
				\bm{A}_{it} & \bm{B}_{it}\\
				\bm{B}'_{it} & \bm{C}_{it}
			\end{pmatrix}\\
			\bm{A}_{it} &= Var(Y_{it})\bm{x}_{it}\bm{x}'_{it} \nonumber\\
			\bm{B}_{it} &= -\nu_{it}Cov(Y_{it}, \log Y_{it}!)\bm{x}_{it}\bm{g}'_{it} \nonumber\\
			\bm{C}_{it} &= \nu_{it}[\nu_{it}Var(\log Y_{it}) - E(\log Y_{it}!) + \log y_{it}!]\bm{g}_{it}\bm{g}'_{it} \nonumber
		\end{align}
		
		When $\log y_{it}! \ll E(\log Y_{it}!)$, the Hessian may be ill-conditioned or even positive-definite. To ensure the robustness, do Fisher scoring, i.e. replace the observed information $-\nabla\nabla_{\bm{\Theta}}f$ by the expected information $E(-\nabla\nabla_{\bm{\Theta}}f)$, so that $\bm{C}_{it} = \nu_{it}^2Var(\log Y_{it}!)\bm{g}_{it}\bm{g}'_{it}$
	\end{appendices}

\end{document}